\documentclass[twocolumn,prl,showpacs,amsmath,amssymb,superscriptaddress]{revtex4-1}

\usepackage[latin1]{inputenc}
\usepackage{graphicx}
\usepackage{dcolumn}
\usepackage{bm}
\begin{document}



\title{Three addressable spin qubits in a molecular single-ion magnet}

\author{M. D. Jenkins}
\affiliation{Instituto de Ciencia de Materiales de Arag\'on, CSIC-Universidad de Zaragoza, Pedro Cerbuna 12, 50009 Zaragoza, Spain}
\affiliation{Dpto. de F\'{\i}sica de la Materia Condensada, Universidad de Zaragoza, Pedro Cerbuna 12, 50009 Zaragoza, Spain}

\author{Y. Duan}
\affiliation{Instituto de Ciencia Molecular (ICMol), Universidad de Valencia, Catedr\'atico Jos\'e Beltr\'an 2, 46980 Paterna, Spain}

\author{B. Diosdado}
\affiliation{Servicio de Difracci\'on de Rayos X y An\'alisis por Fluorescencia, Universidad de Zaragoza, 50009 Zaragoza, Spain}

\author{J. J. Garc\'{\i}a-Ripoll}
\affiliation{Instituto de F\'{\i}sica Fundamental, IFF-CSIC, Serrano 113-bis, 28006 Madrid, Spain}

\author{A. Gaita-Ari\~{n}o}
\affiliation{Instituto de Ciencia Molecular (ICMol), Universidad de Valencia, Catedr\'atico Jos\'e Beltr\'an 2, 46980 Paterna, Spain}

\author{C. Gim\'enez-Saiz}
\affiliation{Instituto de Ciencia Molecular (ICMol), Universidad de Valencia, Catedr\'atico Jos\'e Beltr\'an 2, 46980 Paterna, Spain}

\author{P. J. Alonso}
\email{alonso@unizar.es}
\affiliation{Instituto de Ciencia de Materiales de Arag\'on, CSIC-Universidad de Zaragoza, Pedro Cerbuna 12, 50009 Zaragoza, Spain}
\affiliation{Dpto. de F\'{\i}sica de la Materia Condensada, Universidad de Zaragoza, Pedro Cerbuna 12, 50009 Zaragoza, Spain}

\author{E. Coronado}
\email{eugenio.coronado@uv.es}
\affiliation{Instituto de Ciencia Molecular (ICMol), Universidad de Valencia, Catedr\'atico Jos\'e Beltr\'an 2, 46980 Paterna, Spain}

\author{F. Luis}
\email{fluis@unizar.es}
\affiliation{Instituto de Ciencia de Materiales de Arag\'on, CSIC-Universidad de Zaragoza, Pedro Cerbuna 12, 50009 Zaragoza, Spain}
\affiliation{Dpto. de F\'{\i}sica de la Materia Condensada, Universidad de Zaragoza, Pedro Cerbuna 12, 50009 Zaragoza, Spain}


\date{\today}


\begin{abstract}
We show that several qubits can be integrated in a single magnetic ion, using its internal electronic spin states with energies tuned by a suitably chosen molecular environment. This approach is illustrated with a nearly-isotropic Gd$^{3+}$ ion entrapped in a polyoxometalate molecule. Experiments with microwave technologies, either three dimensional cavities or quantum superconducting circuits, show that this magnetic molecule possesses the number of spin states and the set of coherently addressable transitions connecting these states that are needed to perform a universal three-qubit processor or, equivalently, a $d=8$-level 'qudit'. Our findings open prospects for developing more sophisticated magnetic molecules which can result in more powerful and noise resilient quantum computation schemes.
\end{abstract}

\pacs{75.50.Xx,03.67.Lx,75.45.+j,75.30.Gw}

\maketitle



Molecular nanomagnets have emerged in the last few years as promising candidates to realize qubits, the basic components of future quantum computers \cite{Leuenberger2001,Troiani2005,Ardavan2007,Martinez-Perez2011}. These artificial molecules, designed and synthesized by chemical methods, consist of a magnetic core surrounded by nonmagnetic ligands. They are perfectly monodisperse and remain stable in different material forms, from perfectly ordered crystals to solutions and, in some cases, also when they are deposited onto solid substrates \cite{Mannini2010}. The main sources of magnetic noise, which introduce decoherence, arise from hyperfine couplings to nuclear spins and from dipolar couplings to other electronic spins in their neighborhood \cite{Prokof'ev2000,Morello2006}. These effects can be minimized by isotopical purification and by extreme dilution in a diamagnetic matrix or in appropriate solvents. This strategy has recently led to spin coherence times \cite{Ardavan2007,Bader2014,Zadrozny2015} comparable to those reported for other solid-state spin qubit systems, such as NV centres in diamond and P donors in silicon \cite{Jelezko2004,Pla2012}. An alternative is to identify the qubit states with clock transitions, which are relatively insensitive to magnetic field fluctuations \cite{Shiddiq2016}. However, even if the qubit-qubit interactions are minimized these strategies offer no route to scalability. This paradox between coupling and isolation has severely limited progressing beyond the realization of elemental two-qubit gates with electronic spins \cite{Gaebel2006,Luis2011,Veldhorst2015}.

\begin{figure}[ht]
\resizebox{7.5 cm}{!}{\includegraphics[scale=1]{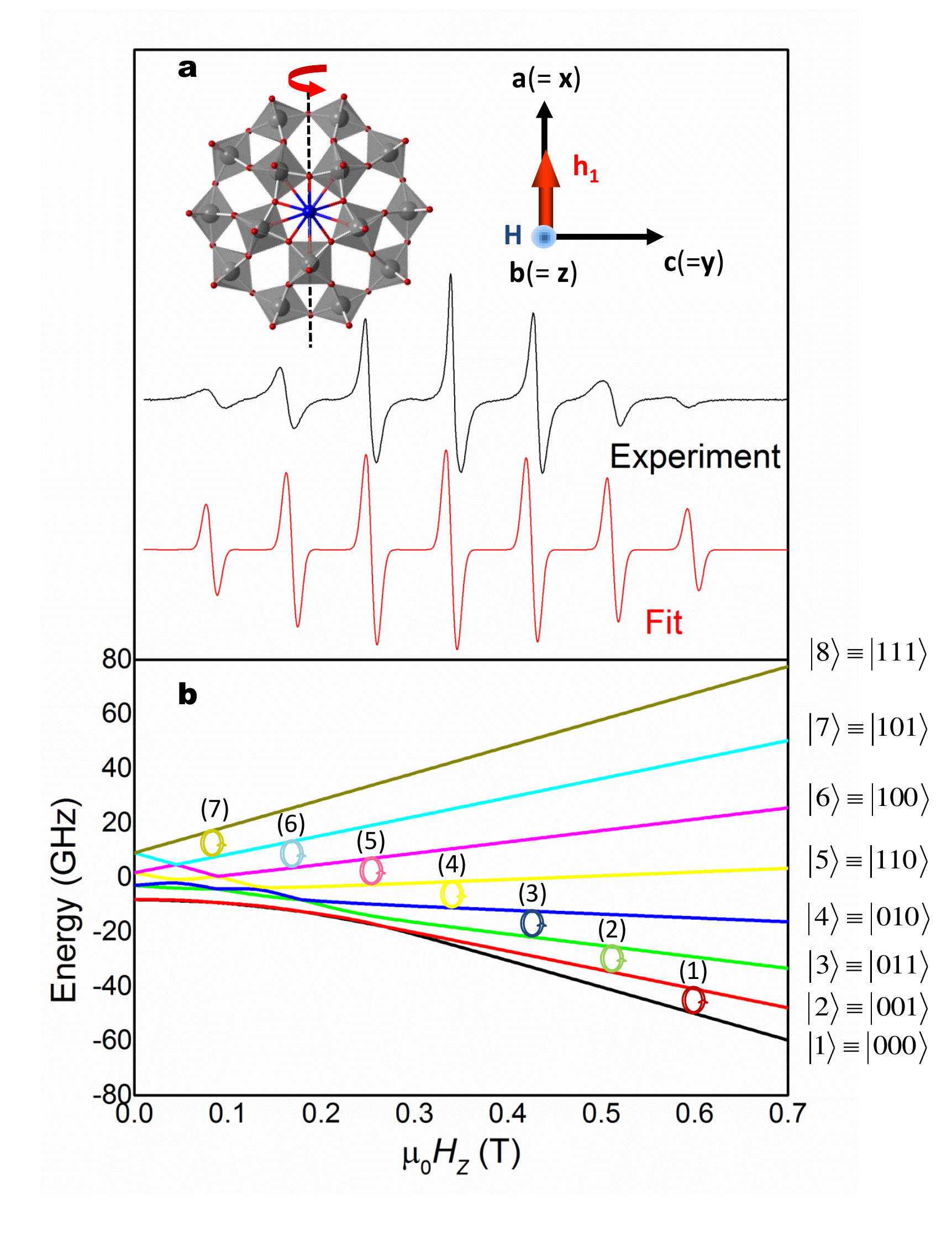}}
\caption{(Color online) (a) X-band cw-EPR spectrum measured at room temperature on a single crystal of Y$_{0.95}$Gd$_{0.05}$W$_{30}$. The crystal was rotated around its $a$ axis to maximize the field splitting of the seven different transitions. The inset shows the structure of each GdW$_{30}$ POM cluster. The arrows show the approximate orientations of the hard ($z$), medium ($x$) and easy ($y$) magnetic axes and the experimental geometry used in the pulsed EPR experiments, where $\vec{H}$ denotes the static magnetic field and $\vec{h}_{1}(t) = 2h_{1} \cos (\omega t)\hat{x}$, with  $\omega/2 \pi  = 9.77$ GHz (X-band), is the microwave magnetic field. (b) Zeeman diagram of the GdW$_{30}$ spin energy levels vs $H_{z}$. A possible assignment for the basis states of a three-qubit quantum processor is shown. Arrows mark the seven addressable transitions between these states.}
\label{fg-level-scheme}
\end{figure}

Here, we explore a different approach: to have multiple qubits coupled in a single atom. Systems having $d = 2^{N}$ internal levels (or 'qudits') can in principle realize $N$ qubits. Lanthanide ions are promising candidates to realize these systems since the multiplet associated with the angular momentum $J$, given by Hund's rules, defines $(2J+1)$ states. Its practical implementation is not straightforward, though, because the level splitting induced by the crystal field around the lanthanide is often so large that only the two lowest lying electronic levels are experimentally accessible \cite{Shiddiq2016,Bertaina2007,Bertaina2009}. Nuclear spin states might still allow the definition of several qubits in lanthanides with a nonzero $I$ \cite{Bertaina2007,Vincent2012}. However, the transitions that connect these states have very different frequencies from those of pure electronic transitions and, furthermore, they depend very weakly on magnetic field and on the molecular environment. For these reasons, it is difficult to tune these qubits and to address them independently from one another.

These difficulties can be overcome by a judicious choice of the lanthanide ion and a proper molecular design of its environment. We here focus on gadolinium that, because of its $4f^{7}$ electronic configuration, has no net orbital moment $L$ but instead possesses the largest spin $S = 7/2$ attainable by any single atom. As illustrated in Fig. \ref{fg-level-scheme}, the $8$ spin states span the Hilbert space of three qubits. This intrinsic property of any free Gd$^{3+}$ ion is, however, not yet sufficient for the purpose of realizing a three-qubit processor. A second necessary ingredient is the existence of a set of coherent transitions able to connect any two arbitrary states. For this, some magnetic anisotropy is required, sufficiently large to enable addressing each transition independently of the others but also sufficiently weak to guarantee that all transition frequencies lie within the range accessible by microwave technologies that form the basis for a scalable quantum architecture \cite{Jenkins2016}. A further advantage of Gd$^{3+}$, in this respect, is that crystal field splittings are between one and two orders of magnitude smaller than those found for other transition metal ions. Furthermore, these splittings strongly depend on the local coordination, which opens subtle possibilities for design \cite{Martinez-Perez2011}.

In this work, the system of choice is a molecular nanomagnet which can be formulated as [Gd(H$_{2}$O)P$_{5}$W$_{30}$O$_{110}$]$^{12-}$. It is hereafter referred to as GdW$_{30}$ and shown in Fig. \ref{fg-level-scheme}. In this molecule, Gd$^{3+}$ is encapsulated inside a "doughnut" shaped polyoxometalate (POM) cluster and coordinated to $10$ oxygen atoms and one apical water. Using potassium as counter-ion, this molecule forms crystals formulated as K$_{12}$Gd(H$_{2}$O)P$_{5}$W$_{30}$O$_{110}\cdot 27.5$H$_{2}$O. The resulting coordination around Gd$^{3+}$ has an unusual planar shape and symmetry close to C$_{5 \rm{v}}$. This leads to an easy-plane spin anisotropy, with a hard magnetic axis along the main molecular axis $z$, and to an overall energy splitting of the $S = 7/2$ multiplet smaller than $1$ K ($20.8$ GHz). The energy level spectrum of GdW$_{30}$ subject to a dc magnetic field $\vec{H}$ can be described by the spin Hamiltonian \cite{Martinez-Perez2011,Baldovi2013}

\begin{equation}
{\cal H} = D \left[ S_{z}^{2}-\frac{1}{3} S \left(S+1 \right) \right] + E \left( S_{x}^{2} - S_{y}^{2} \right) - g\mu_{\rm{B}}\overrightarrow{S}\cdot\overrightarrow{H}
\label{eq-hamiltonian1}
\end{equation}

\noindent where $D = 1281$ MHz and $E = 294$ MHz are second-order magnetic anisotropy constants, $g = 2$ is the gyromagnetic ratio, and the magnetic axes $x$, $y$ and $z$ are oriented as shown in Fig. \ref{fg-level-scheme}a. These parameters lead to the energy spectrum shown in Fig. \ref{fg-level-scheme}b. Almost identical results are obtained for samples with different concentrations of GdW$_{30}$, thus showing that the energy level scheme is an intrinsic property of each isolated molecule. The magnetic anisotropy provides the anharmonicity that is required to address each transition individually but it is also sufficiently weak to make them accessible by conventional X-band ($9.48$ GHz) electron paramagnetic resonance (see Fig. \ref{fg-level-scheme}a). In order to maximize the field splitting between these transitions, thus making them easier to address, $\vec{H}$ was oriented close to the hard magnetic axis of the molecule, $z$. Because of the weak magnetic anisotropy, the eigenstates $|n \rangle$ of the spin Hamiltonian (\ref{eq-hamiltonian1}) are close to spin projections $|m \rangle$  along $\vec{H}$, with $m = -S+(n-1)$, and their energies  $\epsilon_{n} \simeq D[3m^{2} - S(S+1)] - g\mu_{\rm B}H_{z}m$. With the definition of logic states introduced in Fig. \ref{fg-level-scheme}b, single-qubit flips correspond then simply to  $\pm 1$ changes in $m$.

\begin{figure}[t]
\resizebox{6.5 cm}{!}{\includegraphics[scale=1]{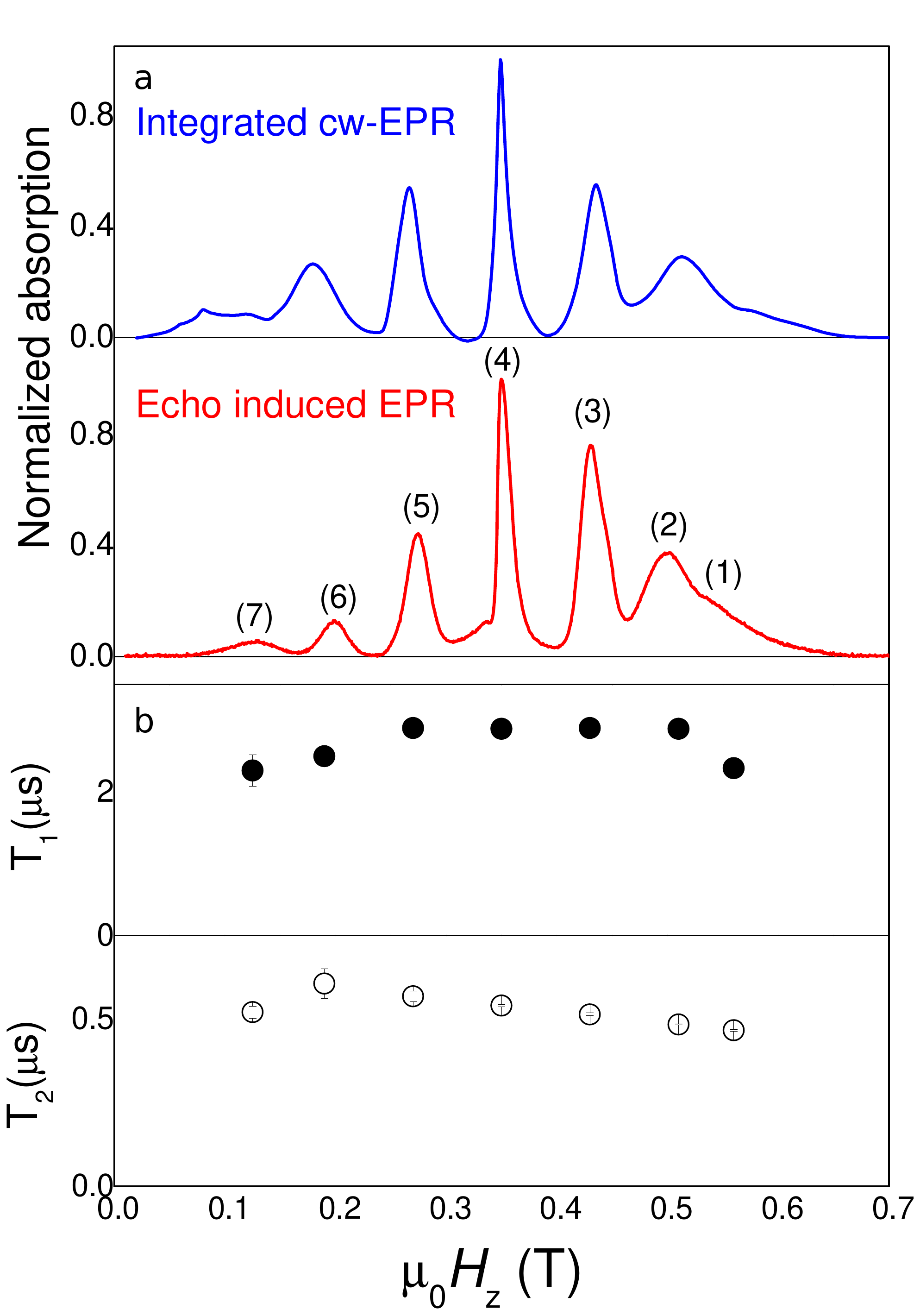}}
\caption{(a) Comparison between the integrated cw-EPR spectrum and the two-pulse echo-induced EPR spectrum measured at $T = 6$ K on a single crystal of Y$_{0.99}$Gd$_{0.01}$W$_{30}$ oriented as shown in Fig. \ref{fg-level-scheme}a. The time interval between  $\pi/2$ and $\pi$ pulses was $\tau = 160$ ns. (b) Spin coherence times ${\rm T}_{2}$ and spin lattice relaxation ${\rm T}_{1}$ times of the seven allowed transitions $n = 7$ to $1$.}
\label{fg-T1T2}
\end{figure}

The spin dynamics has been studied at low temperature ($T = 6$ K) on magnetically diluted single crystals of Y$_{0.99}$Gd$_{0.01}$W$_{30}$, in which $99$ \% of GdW$_{30}$ molecules have been replaced with its diamagnetic YW$_{30}$ equivalent to reduce dipolar interactions. Phase coherence times ${\rm T}_{2}$ have been determined by measuring the spin echo following a Hahn sequence of  $\pi/2$ and $\pi$ pulses resonant with the energy gap  $\Delta_{n} \equiv \epsilon_{n+1} - \epsilon_{n}$ of a given transition ($n = 1$ to $7$) and separated by a varying time interval $\tau$ \cite{supplementary}. The magnetic field dependence of the echo amplitude (Fig. \ref{fg-T1T2}a) shows resonances that closely match those found in the continuous EPR spectrum measured under similar conditions. These results show that all seven transitions can be coherently manipulated. Different transitions show very similar coherence times (Fig. \ref{fg-T1T2}b), ranging approximately between $470$ ns and $600$ ns.

The nuclear spin bath of a GdW$_{30}$ molecule is formed by the protons of water molecules ($100$ \%, $I = 1/2$), and the nuclear spins of P ($100$ \%, $I = 1/2$), K($100$ \%, $I = 3/2$), and W ($14$ \%, $I = 1/2$) isotopes, which couple to the electron spin via dipole-dipole interactions. Recent calculations show that the decoherence induced by these couplings is very weak \cite{Cardona-Serra2016}, with ${\rm T}_{2,n}^{-1} < 1$ kHz or, equivalently, ${\rm T}_{2,n}$ much longer than the values measured experimentally. Spin coherence must, therefore, be limited by pair-flip processes \cite{Morello2006} induced by residual dipolar interactions with other GdW$_{30}$ molecules. The typical dipolar energy of each GdW$_{30}$ molecule diluted in a Y$_{0.99}$Gd$_{0.01}$W$_{30}$ crystal is close to $4.5 \times 10^{-5}$ K $= 0.94$ MHz \cite{Martinez-Perez2011}, thus of the same order of magnitude as the experimental decoherence rates ${\rm T}_{2}^{-1}$. Coherence times of fully isolated GdW$_{30}$ molecules will therefore be limited only by the coupling of their spins to the lattice.

Spin-lattice relaxation times ${\rm T}_{1}$ have been determined by measuring the decay of the echo amplitude after a $3$-pulse sequence \cite{supplementary}. Values of ${\rm T}_{1}$ obtained for the different transitions, shown in Fig. \ref{fg-T1T2}b, range from $2.3$ to $2.8 \mu$s at $T = 6$ K. While not extraordinarily long, the values of ${\rm T}_{2}$ and ${\rm T}_{1}$ nevertheless enable the coherent manipulation of the spin states that form the basis for the three-qubit quantum processor.

Spin nutation experiments were performed by measuring the spin echo amplitude generated by a variable duration excitation pulse ($100$ ns $\leq t_{\rm p} \leq 1 \mu$s) followed, after a fix time interval (typically between $120$ and $200$ ns), by a refocusing  $\pi$-pulse. This method measures the evolution with $t_{\rm p}$ of the transverse spin component $S_{y}$ in the rotating frame. Representative results obtained at the different resonant fields are shown in Fig. \ref{fg-Rabi} and in \cite{supplementary}. For all transitions, $S_{\rm y}$ shows damped coherent oscillations that are well described by the following expression (see \cite{supplementary,Baibekov2014} for a explicit derivation)

\begin{equation}
S_{y,n}(t_{\rm p}) = K_{n} \exp \left( -t_{\rm p}/ \tau_{\rm{R},n} \right) J_{0} \left[ \Omega_{\rm{R},n}(t_{\rm p}-t_{\rm p,0}) \right] S_{y,0}^{\rm{no}}
\label{eq-Rabi}
\end{equation}

\noindent where $K_{n}$ is a constant, $J_{0}$ is the Bessel function of first kind, $\Omega_{\rm{R},n}$ and $1/\tau_{\rm{R},n}$ are, respectively, the frequency and the damping rate of each Rabi oscillation, and the last term is a non-oscillatory component. Values of  $\Omega_{\rm{R},n}$ and $1/\tau_{\rm{R},n}$ are shown in Fig. \ref{fg-Rabi}b.

\begin{figure}[t]
\resizebox{7 cm}{!}{\includegraphics[scale=1]{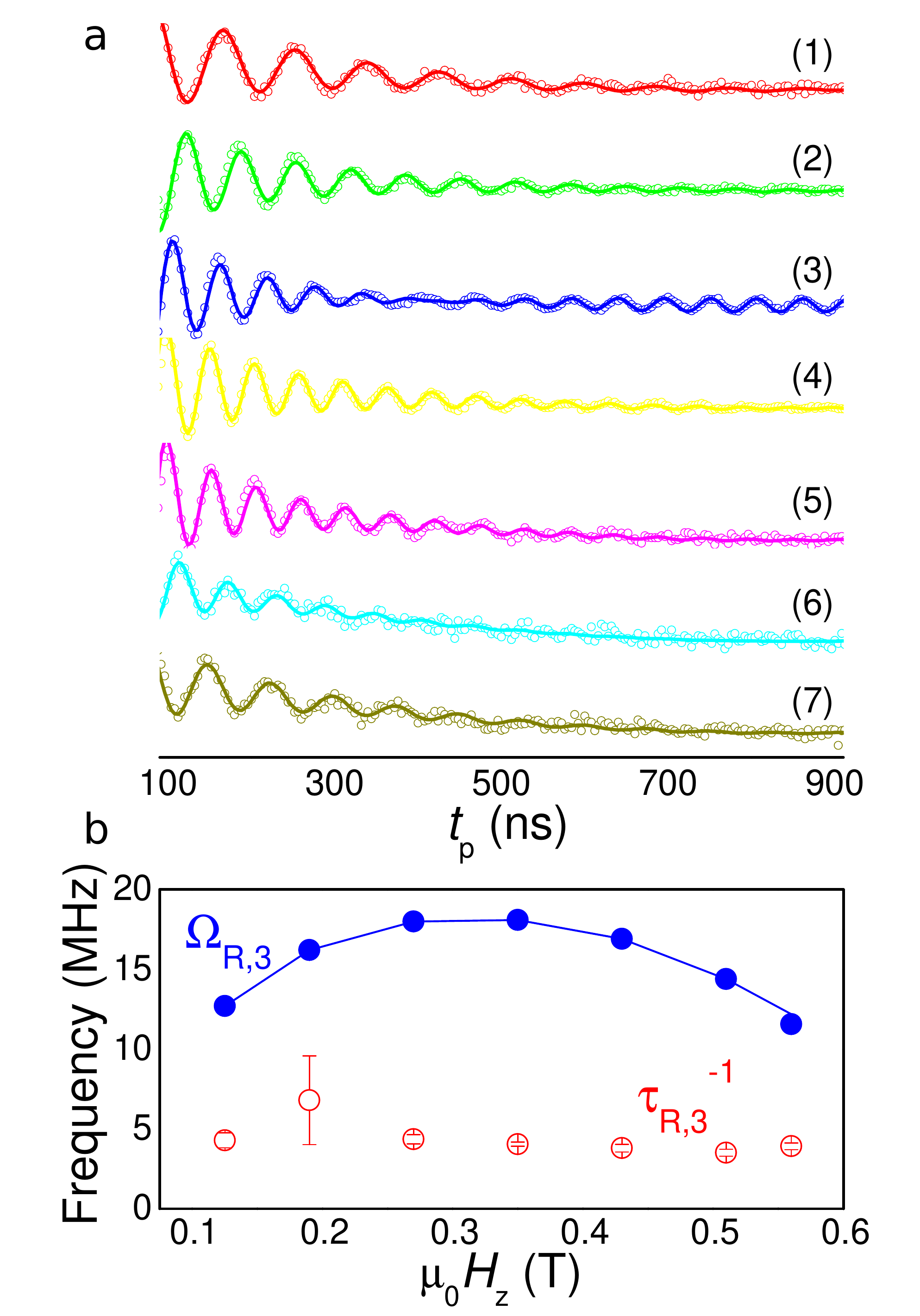}}
\caption{(a) Rabi oscillations for transitions $1-7$ measured on a single crystal of Y$_{0.99}$Gd$_{0.01}$W$_{30}$ oriented as described in Fig. \ref{fg-level-scheme}a. The microwave magnetic field amplitude $\mu_{0}h_{1} = 155(3) \mu$T. The circles are experimental data and the solid lines are least-squares fits based on Eq. \ref{eq-Rabi}. (b) Frequencies $\Omega_{\rm{R},n}$ and damping rates $1/\tau_{\rm{R},n}$ of these oscillations. The solid blue line shows the theoretical $\Omega_{\rm{R},n}$ derived from the spin Hamiltonian (\ref{eq-hamiltonian1}).}
\label{fg-Rabi}
\end{figure}

\begin{figure}[t]
\resizebox{7 cm}{!}{\includegraphics[scale=1]{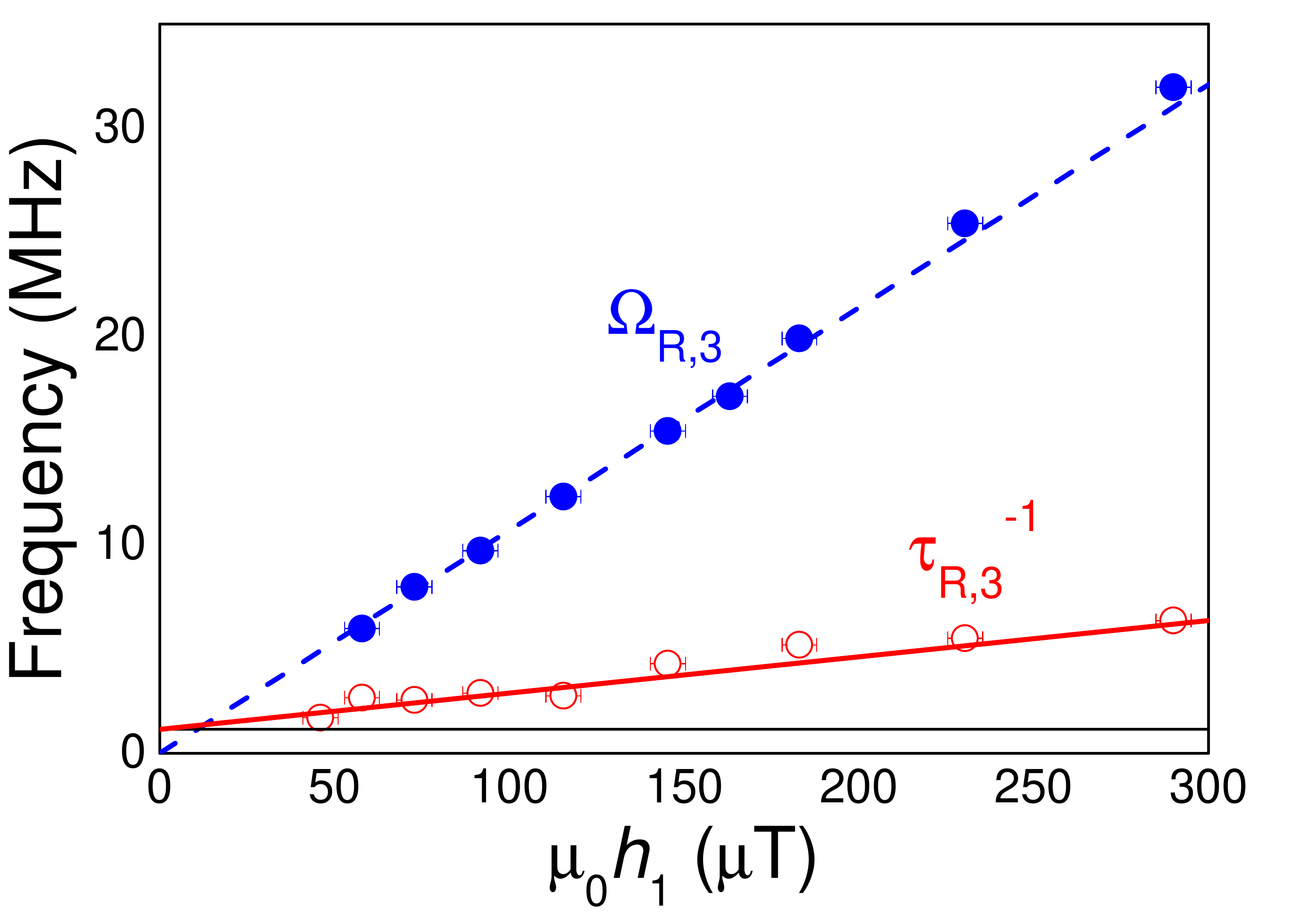}}
\caption{Rabi frequency $\Omega_{\rm{R},3}$ and damping rate $1/\tau_{\rm{R},3}$ of transition $3$, which connects $|011\rangle$ and $|010\rangle$ states, as a function of the microwave magnetic field amplitude $h_{1}$. The dashed blue line is the theoretical prediction derived from the spin Hamiltonian (\ref{eq-hamiltonian1}). The solid red line is a least-squares fit based on $1/\tau_{\rm{R}} = \alpha + \beta \Omega_{\rm{R}}$ (see \cite{supplementary} for a discussion of this equation) while the solid horizontal line gives the low-power limit rate $\alpha = (1/{\rm T}_{1} +1/{\rm T}_{2})/2$ for this transition.}
\label{fg-OmegaR}
\end{figure}

The frequencies of the coherent spin oscillations have been tuned by varying the power of the microwave radiation. The Rabi frequency increases linearly with $h_{1}$ (Fig. 4), in agreement with the theoretical expression  $\Omega_{\rm{R},n} = g \mu_{\rm B}a_{n}h_{1}$, where $a_{n} = \langle n| 2S_{x} | n+1\rangle$  is the transition matrix element for a microwave magnetic field polarized along $x$. Figures \ref{fg-Rabi}b and \ref{fg-OmegaR} show that the experimental Rabi frequencies are, for any $h_{1}$, in very good agreement with those derived from Eq. (\ref{eq-hamiltonian1}). On account of the large spin ($S = 7/2$) of Gd$^{3+}$, the Rabi frequencies for these single qubit flips are larger, for any driving amplitude, than those of qubits realized with $S = 1/2$ spins. The number of observable oscillations does not increase with microwave power in the same manner as $\Omega_{\rm{R},n}$ does, because the oscillation damping rate $1/\tau_{\rm{R},n}$ also increases with $h_{1}$ (Fig. \ref{fg-OmegaR}). This effect limits the quantum quality factors to  $\Omega_{\rm{R},n} \tau_{\rm R} \simeq 5$, lower than the maximum attainable limit $\Omega_{\rm{R},n} \left( \rm{T}_{2}^{-1}+ \rm{T}_{1}^{-1} \right)/2 > 30$. Such behavior is common with lanthanide spin qubits \cite{Bertaina2007,Bertaina2009} and has been ascribed to inhomogeneities and fluctuations of the driving microwave field \cite{Baibekov2011,deRaedt2012}. In the present case, these fluctuations are likely associated with residual dipolar interactions between different molecular spins, which limit the spin coherence times but can easily be minimized by dilution.

As a final test, we check that the allowed transitions form a universal set. In the basis of spin eigenstates $|n\rangle$  the coupling to an external radiation field can be described by the following effective Hamiltonian

\begin{eqnarray}
{\cal H} = & &\sum_{n=1}^{8} \epsilon_{n} |n \rangle\langle n| - 2g\mu_{\rm B} h_{1} \cos (\omega t) \nonumber \\ & \times & \sum_{n=1}^{7} \left( a_{n} |n+1\rangle \langle n| + a^{\ast}_{n} |n \rangle \langle n+1| \right)
\label{eq-hamiltonian2}
\end{eqnarray}

\noindent that connects each level $n$ with its adjacent ones $n \pm 1$. Equation (\ref{eq-hamiltonian2}) contains the generators of a Lie Algebra that includes all unitary operators in this Hilbert space of dimension $d = 8$ and thus it is, by a suitable control of the rf fields $\vec{h}_{1}(t)$, capable of implementing any arbitrary quantum operations on these artificial three-qubit systems. The results can be read out either magnetically or spectroscopically, as each state $|n\rangle$ corresponds to a well-defined spin polarization $|m\rangle$ and it absorbs only photons of frequency $\Delta_{n}/h$.

Universality must be complemented with specific protocols for designing actual quantum operations. As an example, starting from the ground state $|000\rangle$ the sequence of operations $R_{1}(\pi/2)R_{2}(\pi)R_{3}(\pi)R_{4}(\pi)R_{5}(\pi)R_{6}(\pi)R_{7}(\pi)$, where $R_{n}(\theta)$ denotes a rotation by angle $\theta$ tuned at transition $n$, generates a maximally entangled state of the three qubits, namely the GHZ state $|\Psi \rangle = (1/\sqrt{2}) (|000\rangle + |111\rangle)$. Remarkably, this state is entangled in the logic basis but not in the single-ion spin basis, where it corresponds to a quantum superposition of the two states with maximum projections $\pm S$ along $\vec{H}$. This suggests that quantum correlations like those found in composite systems can also occur, and be experimentally tested, in a single magnetic ion.

Summarizing, the previous results show that up to three fully addressable spin qubits can be integrated in a single Gd$^{3+}$ ion and that its characteristic properties can be tuned by the molecular environment. Each GdW$_{30}$ molecule provides then a realization of a 'qudit', that is, of a quantum system with $d = 8$ addressable states. It should be noted that these properties are not exclusive of the Gd$^{3+}$-based polyoxometalate used in the present experiments. The molecular approach used here is particularly attractive for its versatility, since it can produce a vast choice of mononuclear complexes encapsulating single magnetic ions with a spin $S > 1/2$ and with tunable properties \cite{Layfield2015}.

It has been recently proposed that solid-state quantum circuits provide a convenient platform for scaling up quantum computation with molecular nanomagnets \cite{Jenkins2016,Jenkins2013}. We have performed experiments to measure the coupling of GdW$_{30}$ crystals to superconducting coplanar waveguides and resonators. The results \cite{supplementary} confirm that all transitions between different spin states of GdW$_{30}$ are in good speaking terms with these circuits. Besides, its molecular character ensures that they can be taken out of the crystal and placed individually on a surface or in a device \cite{Sheriff2015}.

The integration of several qubits in each molecule will enhance the density of quantum information that can be handled with such schemes and reduce the number of non local gates required to carry out any algorithm \cite{Leuenberger2001,Lanyon2009,Kiktenko2015}. Another interesting application concerns the implementation of quantum correction codes in each molecule \cite{Shor1995,Pirandola2008,Baldovi2015}. Embedding qubits in higher-dimension qudits provides the possibility of restoring its quantum state from some specific errors \cite{Pirandola2008}. A GdW$_{30}$ molecular nanomagnet possesses the minimum dimension $d = 8$ that is required to optimally correct a single amplitude or phase shift error. The operations required to implement the restoring code correspond to either transitions between adjacent spin states or phase shifts \cite{Pirandola2008}, which can be generated by, respectively, transverse or longitudinal magnetic field pulses. The extension of this rational design to molecules containing two or three weakly coupled Gd$^{3+}$ ions with distinct coordinations has also been shown to be feasible \cite{Aguila2014}. Scaling up to $6$ or $9$ addressable qubits within a molecule, which would then be able to optimally correct any arbitrary error \cite{Shor1995}, is, therefore, within reach. In conclusion, the new strategy reported here contributes to the development of more efficient and noise-resilient quantum information schemes based on molecular nanomagnets.


\begin{acknowledgments}

The authors acknowledge useful discussions with Seiji Miyashita and Jos\'e Luis Garc\'{\i}a-Palacios. The research reported here was supported by the Spanish MINECO (grants MAT2015-68204-R, CTQ2015-64486-R, MAT-2014-56143-R, CTQ2014-52758-P, FIS2015-70856-P, and Excellence Unit Mar\'{\i}a de Maeztu MDM-2015-0538), the European Union (ERC grants SPINMOL and DECRESIM, and COST 15128 Molecular Spintronics project), the Gobierno de Aragón (grants E98-MOLCHIP and E33), Comunidad de Madrid (Research Network QUITEMAD+) and the Generalidad Valenciana (Prometeo and ISIC-Nano Programs of Excellence). A.G.-A. thanks the Spanish MINECO for a Ram\'on y Cajal Fellowship.
\end{acknowledgments}




\begin{thebibliography}{10}

\bibitem{Leuenberger2001} M. N. Leuenberger and D. Loss, Nature {\bf 410}, 789 (2001).

\bibitem{Troiani2005}	F. Troiani, A. Ghirri, M. Affronte, S. Carretta, P. Santini, G. Amoretti, S. Piligkos, G. Timco, R. E. P. Winpenny, Phys. Rev. Lett. {\bf 94}, 207208 (1-4) (2005).

\bibitem{Ardavan2007}	A. Ardavan, O. Rival, J. J. L. Morton, S. J. Blundell, A. M. Tyryshkin, G. A. Timco, R. E. P. Winpenny, Phys. Rev. Lett. {\bf 98}, 057201 (1-4) (2007).

\bibitem{Martinez-Perez2011} M. J. Mart\'{\i}nez-P\'erez, S. Cardona-Serra, C. Schlegel, F. Moro, P. J. Alonso, H. Prima-Garc\'{\i}a, J. M. Clemente-Juan, M. Evangelisti, A. Gaita-Ari\~{n}o, J. Ses\'e, J. van Slageren, E. Coronado and F. Luis, Phys Rev. Lett. {\bf 108}, 247213 (1-5) (2012).

\bibitem{Mannini2010} M. Mannini, F. Pineider, C. Danieli, F. Totti, L. Sorace, Ph. Sainctavit, M. A. Arrio, E. Otero, L. Joly, J. C. Cezar, A. Cornia, and R. Sessoli, Nature {\bf 468}, 417 (2010).

\bibitem{Prokof'ev2000} N. V. Prokof'ev and P. C. E. Stamp, Rep. Prog. Phys. {\bf 63}, 669 (2000).

\bibitem{Morello2006}	A. Morello, P. C. E. Stamp and I. S. Tupitsyn, Phys. Rev. Lett. {\bf 97}, 207206 (1-4) (2006).

\bibitem{Bader2014}	K. Bader, D. Dengler, S. Lenz, B. Endeward, S.-D. Jiang, P. Neugebauer, J. van Slageren, Nature Commun. {\bf 5}, 5304 (1-5) (2014).

\bibitem{Zadrozny2015}	J. M. Zadrozny, J. Niklas, O. G. Poluektov, D. E. Freedman, ACS Cent. Sci. {\bf 1}, 488 (2015).


\bibitem{Jelezko2004} F. Jelezko, T. Gaebel, I. Popa, A. Gruber and J. Wrachtrup, Phys. Rev. Lett. {\bf 92}, 076401 (1-4) (2004).

\bibitem{Pla2012} J. J. Pla, K. Y. Tan, J. P. Dehollain, W. H. Lim, J. J. L. Morton, D. N. Jamieson, A. S. Dzurak and A. Morello, Nature {\bf 489}, 541 (2012).
    
\bibitem{Shiddiq2016} M. Shiddiq, D. Komijani, Y. Duan, A. Gaita-Ari\~{n}o, E. Coronado and S. Hill, Nature {\bf 531}, 348 (2016).

\bibitem{Gaebel2006} T. Gaebel, M. Domhan, I. Popa, C. Wittmann, P. Neumann, F. Jelezko, J. R. Rabeau, N. Stavrias, A. D. Greentree, S. Prawer, J. Meijer, J. Twamley, P. R. Hemmer and J. Wrachtrup, Nature Phys. {\bf 2}, 408 (2006).

\bibitem{Luis2011}	F. Luis, A. Repollés, M. J. Mart\'{\i}nez-P\'erez, D. Aguil\`{a}, O. Roubeau, D. Zueco, P. J. Alonso, M. Evangelisti, A. Cam\'on, J. Ses\'e, L. A. Barrios, G. Arom\'{\i}, Phys. Rev. Lett. {\bf 107}, 117203 (1-5) (2011).

\bibitem{Veldhorst2015}	M. Veldhorst, C. H. Yang, J. C. C. Hwang, W. Huang, J. Dehollain, J. T. Muhonen, S. Simmons, A. Laucht, F. E. Hudson, K. M. Itoh, A. Morello and A. S. Dzurak, Nature {\bf 526}, 410 (2015).

\bibitem{Bertaina2007}	S. Bertaina, S. Gambarelli, A. Tkachuk, I. N. Kurkin, B. Malkin, A. Stepanov and B. Barbara, Nature Nanotech. {\bf 2}, 39 (2007).

\bibitem{Bertaina2009}	S. Bertaina, J. H. Shim, S. Gambarelli, B. Z. Malkin and B. Barbara, Phys. Rev. Lett. {\bf 103}, 226402 (1-4) (2009).

\bibitem{Vincent2012} R. Vincent, S. Klyatskaya, M. Ruben, W. Wernsdorfer and F. Balestro,
    Nature {\bf 488}, 357 (2012).

\bibitem{Jenkins2016} M. D. Jenkins, D. Zueco, O. Roubeau, G. Arom\'{\i}, J. Majer and F. Luis, Dalton Trans. (2016), DOI: 10.1039/c6dt02664h.

\bibitem{Baldovi2013}	J. J. Baldov\'{\i}, S. Cardona-Serra, J. M. Clemente-Juan, E. Coronado, A. Gaita-Ari\~{n}o and H. Prima-Garc\'{\i}a, Chem. Commun. {\bf 49}, 8922 (2013).

\bibitem{supplementary} See accompanying supplementary information.

\bibitem{Cardona-Serra2016}	S. Cardona-Serra, L. Escalera-Moreno, J. J. Baldov\'{\i}, A. Gaita-Ari\~{n}o, J. M. Clemente-Juan and E. Coronado, J. Comp. Chem. {\bf 37}, 1238 (2016).

\bibitem{Baibekov2014}	E. I. Baibekov, Appl. Magn. Reson. {\bf 45}, 1289 (2014).

\bibitem{deRaedt2012} H. De Raedt, B. Barbara, S. Miyashita, K. Michielsen, S. Bertaina and S.
    Gambarelli, Phys. Rev. B {\bf 85}, 014408 (1-17) (2012).

\bibitem{Baibekov2011}	E. I. Baibekov, JETP letters {\bf 93}, 292 (2011).

\bibitem{Layfield2015}	{\em Lanthanides and Actinides in Molecular Magnetism}, R. Layfield and M. Murugesu Eds. Wiley-VCH (2015).

\bibitem{Jenkins2013} M. Jenkins, T. H\"{u}mmer, M. J. Mart\'{\i}nez-P\'erez, J. J.  Garc\'{\i}a-Ripoll, D. Zueco, and F. Luis, New J. Phys. {\bf 15}, 095007 (1-22) (2013).

\bibitem{Sheriff2015}	S. Sherif, {\em et al}, Nanotechnology {\bf 26}, 291001 (1-6) (2015).

\bibitem{Lanyon2009} B. P. Lanyon, M. Barbieri, M. P. Alemida, Th. Jenewein, T. C. Ralph, K. J. Resch, G. J. Pryde, J. L. O'Brien, A. Gilchrist and A. G. White, Nature Physics {\bf 5}, 134 (2009).

\bibitem{Kiktenko2015}	E. O. Kiktenko, A. K. Fedorov, A. A. Strakhov and V. I. Man'ko, Phys. Lett. A {\bf 379}, 1409 (2015).

\bibitem{Shor1995}	P. Shor, Phys. Rev. A {\bf 52}, R2493 (1995).

\bibitem{Pirandola2008}	S. Pirandola, S. Mancini, S. L. Braunstein, and D. Vitali, Phys. Rev. A {\bf 77}, 032309 (1-9) (2008).

\bibitem{Baldovi2015}	J. J. Baldov\'{\i}, S. Cardona-Serra, J. M. Clemente-Juan, L. Escalera-Moreno, A. Gaita-Ari\~{n}o and G. M\'{\i}nguez-Espallargas, EPL {\bf110}, 33001 (1-5) (2015).

\bibitem{Aguila2014} D. Aguil\`{a}, L. A. Barrios, V. Velasco, O. Roubeau, A. Repoll\'es, P. J. Alonso, J. Ses\'e, S. J. Teat, F. Luis and G. Arom\'{\i}, J. Am. Chem. Soc. {\bf 136}, 14215 (2014).


\end{thebibliography}
\end{document}